\documentclass[useAMS, usenatbib]{mn2e}

\usepackage{deluxetable}
\usepackage{epsfig} 
\usepackage{subfigure}
\usepackage{comment}
\usepackage{graphicx}
\usepackage{epstopdf}
\usepackage{lscape}
\usepackage{amsmath}
\usepackage{amssymb}
\usepackage{color}
\usepackage{hyperref}

\newcommand{\apj}{ApJ}

\newcommand{\apjl}{ApJL}
\newcommand{\apjs}{ApJS}
\newcommand{\mnras}{MNRAS}

\newcommand{\Msun}{M_{\odot}}

\newcommand{\beq}{\begin{eqnarray}}
\newcommand{\eeq}{\end{eqnarray}}

\newcommand{\vonetwo}{v_{12}}
\newcommand{\lcdm}{\Lambda\rm{CDM}}
\newcommand{\kms}{\,\rm{km}\,\rm{s}^{-1}}
\newcommand{\monetwo}{M_1,M_2}

\newcommand{\RS}{{\sc Rockstar}}

\newcommand{\PRSMB}{4.58\times10^{-4}}        
\newcommand{\PFOFMB}{1.35\times10^{-5}}      
\newcommand{\nRSMB}{1.23\times10^{-9}}         
\newcommand{\nFOFMB}{3.45\times10^{-11}}     
\newcommand{\BLPRSMB}{1.36\times10^{-3}}    
\newcommand{\BLPFOFMB}{7.38\times10^{-6}}  
\newcommand{\BLnRSMB}{1.52\times10^{-10}}     

\newcommand{\PRSLF}{7.57\times10^{-4}}        
\newcommand{\PFOFLF}{2.89\times10^{-5}}      
\newcommand{\nRSLF}{2.04\times10^{-9}}         
\newcommand{\nFOFLF}{7.41\times10^{-11}}     
\newcommand{\BLPRSLF}{5.84\times10^{-2}}    
\newcommand{\BLPFOFLF}{9.41\times10^{-6}}  
\newcommand{\BLnRSLF}{7.92\times10^{-11}}    

\newcommand{\BOXONE}{L1607 }
\newcommand{\BOXTWO}{L3214 }
\newcommand{\BOXTHREE}{L6249 }

\newcommand{\sampleOne}{{\it high-$\vonetwo$ }}
\newcommand{\sampleTwo}{{\it massive, high-$\vonetwo$ }}
\newcommand{\SampleOne}{{\it High-$\vonetwo$ }}
\newcommand{\SampleTwo}{{\it Massive, high-$\vonetwo$ }}

\title[The Bullet Cluster in $\Lambda$ Cold Dark Matter simulations]{The rise and fall of a challenger: the Bullet Cluster in {\boldmath $\Lambda$} cold dark matter simulations}

\author[Thompson, Dav{\'e}, \& Nagamine]{Robert Thompson$^{1,2}$,  Romeel Dav{\'e}$^{1,3,4}$, Kentaro Nagamine$^{5,6}$\\
$^1$ University of the Western Cape, Bellville, Cape Town 7535, South Africa\\
$^2$ Astronomy Department, University of Arizona, Tucson, AZ 85721, USA\\
$^3$ South African Astronomical Observatory, Observatory, Cape Town 7925, South Africa\\
$^4$ African Institute for Mathematical Sciences, Muizenberg, Cape Town 7945, South Africa\\
$^5$ Department of Earth and Space Science, Graduate School of Science, Osaka University, \\
1-1 Machikaneyama-cho, Toyonaka, Osaka 560-0043, Japan \\
$^6$ Department of Physics and Astronomy, University of Nevada, Las Vegas, 4505 S. Maryland Pkwy, Las Vegas, NV 89154-4002, USA
}

\begin{document}

\maketitle 


 \begin{abstract}
The Bullet Cluster has provided some of the best evidence for the $\Lambda$ cold dark matter ($\lcdm$) model
via direct empirical proof of the existence of collisionless dark matter, while posing a serious challenge owing 
to the unusually high inferred pairwise velocities of its progenitor
clusters.  Here we investigate the probability of finding such a
high-velocity pair in large-volume N-body simulations, particularly focusing on
differences between halo finding algorithms.  We find that algorithms
that do not account for the kinematics of infalling groups yield
vastly different statistics and probabilities.  When employing the
\RS\ halo finder that considers particle velocities, we find
numerous Bullet-like pair candidates that closely match not only
the high pairwise velocity, 
but also the mass, mass ratio,
separation distance, and collision angle of the initial conditions
that have been shown to produce the Bullet Cluster in non-cosmological
hydrodynamic simulations.  
The probability of finding a high pairwise velocity pair
among haloes with $M_{\rm halo}\geq10^{14} \Msun$ is $4.6\times 10^{-4}$
using \RS, while it is $\approx 34\times$ lower using a
friends-of-friends (FOF) based approach as in previous studies.
This is because the typical spatial extent of Bullet progenitors is
such that FOF tends to group them
into a single halo despite clearly distinct kinematics.
Further requiring an appropriately high average mass among the two
progenitors, we find the comoving number density of potential Bullet-like candidates to
be on the order of 
$\approx10^{-10} \,{\rm Mpc}^{-3}$.
Our findings suggest that $\lcdm$ straightforwardly produces massive,
high relative velocity halo pairs analogous to Bullet Cluster
progenitors, and hence the Bullet Cluster does not present a challenge
to the $\lcdm$ model.

\end{abstract}

\begin{keywords}
method : N-body simulations --- galaxies : evolution --- galaxies : formation --- galaxies: clusters --- cosmology : theory --- cosmology : dark matter
\end{keywords}

\section{Introduction}
 \label{sec:intro}
  
Observations of merging massive clusters such as the Bullet Cluster (1E0657-56)
provide a unique opportunity to test the $\lcdm$ paradigm.  This
particular object consists of two massive clusters that have recently
passed through one another and are separated by $\simeq0.72\rm{Mpc}$
on the sky at an observed redshift of $z=0.296$
\citep{Clowe04,Clowe06,Bradac06}.  This system is relatively unique
due to the collision trajectory being almost perpendicular to our
line of sight.  Both clusters are also quite massive and hence rare, with
M$_{\rm{parent}}\simeq1.5\times10^{15}\Msun$ \&
M$_{\rm{bullet}}\simeq1.5\times10^{14}\Msun$ \citep{Clowe04,Clowe06,Bradac06}.
{\it Chandra} X-ray observations revealed that the primary
baryonic component has been stripped away from the primary mass
component (identified via weak lensing) in the collision, and resides 
between the two massive
clusters in the form of hot X-ray emitting gas \citep{Mark06}.  This
evidence provided direct empirical proof for the existence of 
collisionless and mass-dominant dark matter \citep{Clowe06}.

Shock features in the gas have been used to infer the velocity of
the bow shock preceding the `bullet'
\citep[$v_{\rm{shock}}=4740^{+710}_{-550}\kms$;][]{Mark06}, which
was initially assumed to be approximately the infall velocity of the 
`bullet' itself.  Through the use of non-cosmological hydrodynamic
simulations, several groups have shown that this is not necessarily
the case \citep{Milo07,Springel07,Mast08,Lage14}.  
Halo pair (indicated by the subscript `$12$')
initial
configurations varied with separation distances ($d_{12}$) ranging from $\sim 3.4 - 5$\,Mpc,
$M_{\rm{parent}}$ ($M_1$) from $7.13\times10^{14} - 1.91\times10^{15}\Msun$, $M_{\rm{bullet}}$ ($M_2$) from $1.14\times10^{14} - 2.59\times10^{14}\Msun$,
and pairwise velocities ($\vonetwo$) ranging from $2057-3980\kms$ at $z \simeq 0.5$.  
\citet{Mast08} in particular, set the initial halo requirements to be 
M$_1\simeq7.13\times10^{14}\Msun$, M$_2\simeq1.14\times10^{14}\Msun$, and
$\vonetwo\simeq3000\kms$ at a separation distance of $d_{12}\simeq 5$\,Mpc.
More recent work by \citet{Lage14b} has revised these values to 
M$_1\simeq1.91\times10^{15}\Msun$, M$_2\simeq2.59\times10^{14}\Msun$, and
$\vonetwo\simeq2799\kms$ at a separation distance of $d_{12}\simeq 2.8$\,Mpc.

Reproducing such a massive, close, and high-$\vonetwo$ merging pair
in large N-body cosmological simulations has proven to be very
challenging \citep{Hayashi06,Lee10,Thompson12,Watson14,Bouillot14}, potentially
suggesting that the canonical
$\lcdm$ model with Gaussian perturbations is inconsistent with
the observed Bullet Cluster.  Improving upon the work of \citet{Lee10},
\citet{Thompson12} calculated the probability of finding a halo pair 
with $\vonetwo\geq3000\kms$ among all halo pairs with
$\monetwo\geq10^{14}\Msun$
 and $d_{12}\leq10\,\rm{Mpc}$ to be $\rm{P}=2.8\times10^{-8}$.
Extrapolating their cumulative $\vonetwo$ curve,
they estimated that one would need a box size of at least
$\simeq(6.25\, \rm{Gpc})^3$
to produce one Bullet-like pair.

\citet{Bouillot14} argue that the simulations of \citet{Lee10} and \citet{Thompson12}
were too limited in volume 
(($4.3\,\rm{Gpc})^3$ and ($2.8\,\rm{Gpc})^3$ respectively)
to properly characterise the tail of the $\vonetwo$ distribution.
They estimated the probability of finding a Bullet-like cluster in
a 
($29\,\rm{Gpc})^3$ 
simulation to be
P($\vonetwo>3000\kms$)$=6.4\times10^{-6}$, which is two orders of
magnitude larger than estimates by \citet{Thompson12}.  However,
even with an improved probability in such a large volume,
\citet{Bouillot14} did not find any halo pairs matching the initial
configurations required to reproduce the observed properties of
1E0657-56 \citep{Mast08,Lage14b} .

The simulations analysed by \citet{Lee10}, \citet{Thompson12}, and \citet{Bouillot14}
have one crucial aspect in common: each group used a variant of the
friends-of-friends (FOF) algorithm \citep[e.g.,][]{Davis85} to identify and
group their dark matter particles into haloes.  It is known that FOF
tends to `over-group' the dark matter haloes when the resolution of
the simulation is not adequate.  It is often the case that a trace
amount of particles bridge the two haloes, resulting in them being
identified as a single dumbbell-shaped group.  When the overlap between
the two haloes is more significant, FOF has no way of separating
them into two components.  

In this paper, we demonstrate that in the context of searching for
a close, massive, high-$\vonetwo$ pair, one cannot accurately identify haloes
based solely on the spatial distribution of particles, as FOF does.
To properly separate and identify substructures, we must also
consider the particle velocities.  The recently developed \RS\
halo finder~\citep{ROCKSTAR} provides a way to do so.  We use
\RS\ to calculate more robust statistics and probabilities for
finding 
close, massive, high-$\vonetwo$
halo pairs in a large cosmological N-body simulation.
We find much greater numbers of such pairs
than in
previous works, and moreover they reasonably match the required
initial configurations of \citet{Mast08} and \citet{Lage14b} in mass, mass
ratio, $d_{12}$, collision angle, and $\vonetwo$.

This paper is organized as follows: in Section~\ref{sec:simulations}
we detail our simulations.  Section~\ref{sec:halofinders} details
the halo finding algorithms.  We present our results in
Section~\ref{sec:results}.   Section~\ref{sec:conclusions} contains
concluding remarks and discussion.

\section{Simulations}
 \label{sec:simulations}
For our simulations we use the {\small GADGET-3} code \citep{Springel05},
which simulates large N-body systems by means of calculating
gravitational interactions with a hierarchical multipole expansion.
It uses a particle-mesh method for long-range forces and a tree
method for short-range forces.

Initial conditions are initialized at $z=99$ using {\small N-GenIC}\footnote{\url{http://www.mpa-garching.mpg.de/gadget/}}.
We assume cosmological parameters
consistent with constraints from the WMAP \citep{WMAP9} \& Planck
\citep{Planck13} results, namely $\Omega_{m}=0.3,
\Omega_{\Lambda}=0.7,H_0=70,\sigma_8=0.8,n_s=0.96$.  Our largest
simulation employs $1600^3$ collision-less dark matter (DM) particles
in a 
$(6.4\,\rm{Gpc})^{3}$
volume with an effective force
resolution of 
$\epsilon=160.7\,\rm{kpc}$
(i.e., comoving gravitational softening length).  
Two simulations with smaller
volumes 
(($3.2\,\rm{Gpc})^3$ and ($1.6\,\rm{Gpc})^3$)
and particle counts
($800^3$ and $400^3$) were ran with the same force resolution to
test how the simulation volume affects our results.  A summary of
the simulations used in this study can be found in Table~\ref{table:sims}.

\begin{deluxetable}{ccrcc}
\tablecolumns{5}
\tablewidth{0pc}
\tablecaption{Simulations \, \, \, \, \, \, \, \, \, \, \, \, \, \, \, \, \, \, \, \, \, \, \, \, \, \, \, \, \, \, \, \, \, \, \, \, \, \, \, \, \, \, \,  \, \, \, \, \, \, \, \, \, \, \, \, \, }
\tablehead{\multicolumn{1}{c}{Run} &
	\colhead{Box Size} & 
	\colhead{Particle } &
	\colhead{M$_{dm}$} &
	\colhead{$\epsilon$} \\
	\colhead{Name} &
	\colhead{[Mpc]} &
	\colhead{ Count } &
	\colhead{[$\Msun$]} &
	\colhead{[kpc]}
	}
\startdata
\BOXTHREE & 6429 & $1600^3$ & $2.64\times10^{12}$ & 160.7 \\
\BOXTWO & 3214 & $800^3$  & $2.64\times10^{12}$ & 160.7 \\
\BOXONE & 1607 & $400^3$ & $2.64\times10^{12}$   & 160.7 \\
\enddata
\vspace{-0.5cm}
\label{table:sims}
\tablecomments{Summary of simulations used in this paper.  M$_{dm}$ is the mass of each dark matter particle, and $\epsilon$ is the comoving gravitational softening length.
We have incorporated the effect of the Hubble parameter `$h$' in the numbers shown in the table and throughout this paper.
}
\end{deluxetable}

\section{Halo Finders}
 \label{sec:halofinders}
Identifying dark matter haloes as groups of particles within simulation data
is a challenging affair, and there are numerous codes
with different feature sets employing different algorithms.
\citet{MAD11} compared a number of halo finders in both cosmological
and idealized scenarios.  Overall, they found most to be in agreement
with one another, with only subtle variations among the results.
But in detail and for specific types of systems, the differences
can be substantial.
Here we employ two popular group finding algorithms to group
dark matter particles into haloes: a friends-of-friends algorithm,
and a six-dimensional phase-space halo finder.

The FOF algorithm used in this study is a simplified version of the
parallel friends-of-friends group finder {\small SUBFIND}
\citep{Springel01}
with the post-processing {\small SUBFIND} algorithm disabled.
The code groups the particles into DM haloes
if their positions lie within a specified linking length.  This
linking length is a fraction of the initial mean inter-particle
separation, for which we adopt a standard value of $b=0.15$
\citep{More11}.  Additional groupings with $b=0.20$ were performed,
whose results are briefly discussed in Section~\ref{sec:pdf}.

We also use a six-dimensional phase-space algorithm called 
\RS\ \citep[hereafter RS;][]{ROCKSTAR}, which is based on an
adaptive hierarchal refinement of friends-of-friends groups in both
positional and velocity space.  
RS initially divides the simulation into 3D friends-of-friends groups using a $b=0.28$.
The velocity and positions of each halo are then normalized by the group position and velocity dispersions, which gives rise to a natural phase-space metric.
Next, a phase-space linking length is adaptively chosen such that 70\% of the group's particles are linked together in subgroups; the normalization process is then repeated for each subgroup.
Once all levels of substructure are found, seed haloes are placed at the lowest substructure levels and particles are assigned hierarchically to the closest seed halo in phase space.
Finally, unbound particles are removed from the group.
This allows RS to more accurately
identify substructure while maintaining accurate recovery of halo
properties \citep[see][for further details]{MAD11}.

For most situations, FOF determines halo properties to $10\%$
accuracy \citep{MAD11}.  The algorithm however, is not without
weaknesses.  In major mergers or when sub-haloes lie close to the
centers of their host haloes, the density contrast is not strong
enough to distinguish which particles belong to which halo.  If the
two haloes have some relative motion, six-dimensional halo finders
(such as RS) can additionally use particle velocity information to
determine halo membership \citep{ROCKSTAR}.

\section{Results}
\label{sec:results}

To search for Bullet Cluster-like halo pair progenitors, 
we examine our simulations at $z=0.489$ 
to identify systems with the required initial configurations
of both \citet[][hereafter MB08]{Mast08} and \citet[][hereafter LF14]{Lage14b}.

\subsection{Halo Mass Function}
 \label{sec:massfunc}

\begin{figure}
\includegraphics[scale=0.45]{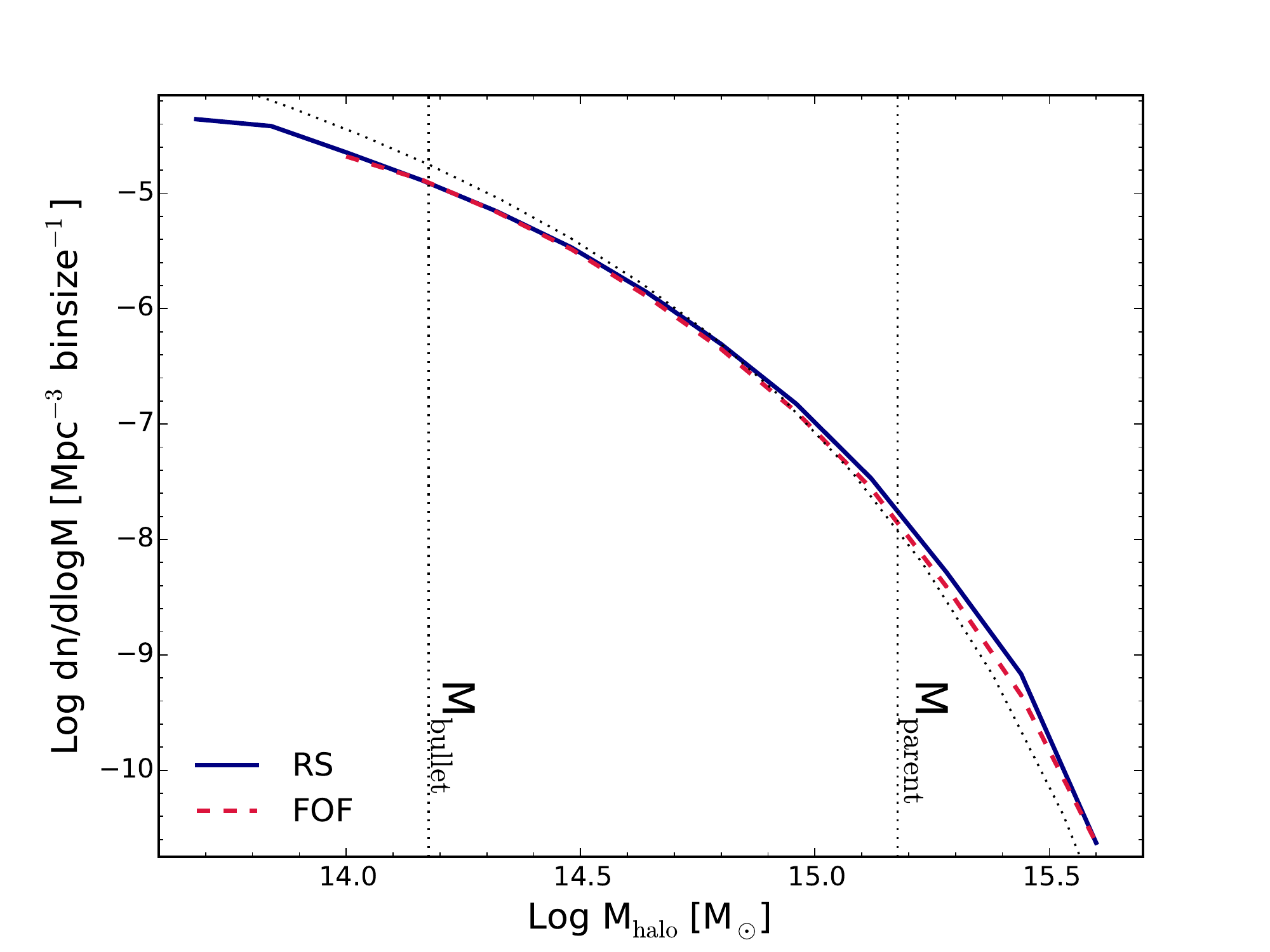}
\caption{DM halo mass function at $z=0.489$ for RS \& FOF groupings.  The vertical black dotted lines represent the mass of the observed bullet and parent respectively 
\citep{Clowe04,Clowe06,Bradac06}.  
The dashed line represents the theoretical DM halo mass function from \citet{ST99}.}
\label{fig:massfunc}
\end{figure}

To check the validity of our DM halo identification,  we examine
the DM halo mass function in Figure~\ref{fig:massfunc}.  The mass
functions of the two halo finders match remarkably well.  This
agreement is not surprising, because the virial masses calculated
by RS include all substructure and should be comparable to the FOF
halo masses.  We truncate the FOF halo mass function at 32 particles,
but we show the RS mass function below this in order to visualize
the level of incompleteness owing to poor numerical resolution at
low halo masses; we will only be concerned with haloes 
$\geq10^{14}\Msun$, above which the mass function is not limited by
our resolution, and in this regime there is little difference in
the mass function between the two codes.

The black dashed line in Figure~\ref{fig:massfunc} is the \citet{ST99}
DM halo mass function at $z=0.489$.  Our simulations with both
groupings slightly underpredict with regards to analytic theory at
the low-mass end, and slightly over predict at the high-mass end.
Many studies have shown that theoretical models such as \citet{ST99}
do not always agree with simulations since they do not capture the
entire complexity of halo formation \citep[i.e.
][]{Jenkins01,Tinker08,Robertson09,Courtin11}.  The important point
here however, is that both halo identification codes agree well
with each other at M$_{\rm halo}\geq1\times10^{14}\Msun$, which
corresponds to a halo with approximately 40 DM particles.

\subsection{Average pair mass and pairwise velocities}
 \label{sec:v12}
\begin{figure}
\includegraphics[scale=0.33]{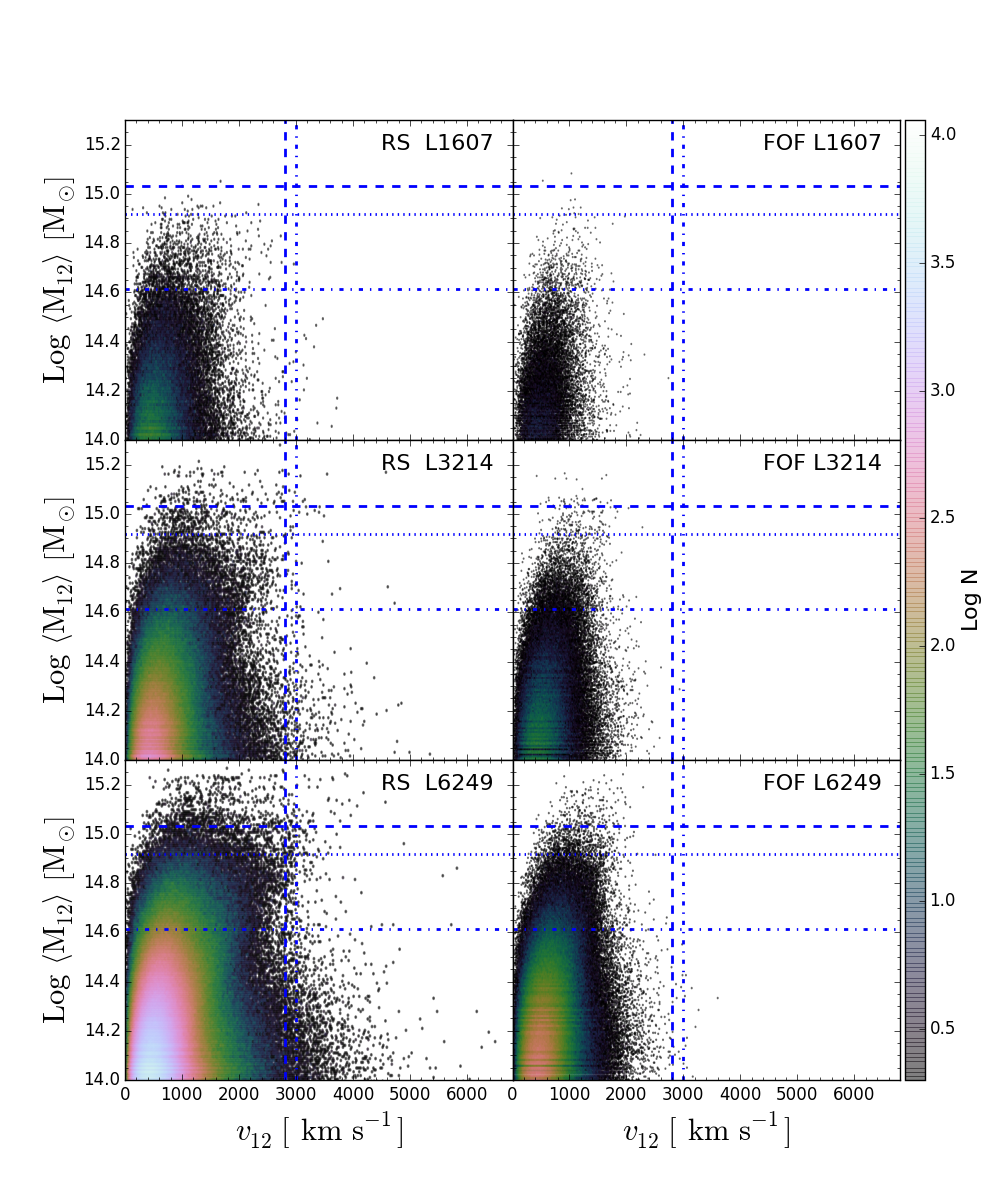}
\caption{
Average halo pair mass $\langle M_{12}\rangle$ as a function of halo pairwise velocity $\vonetwo$.  
Color corresponds to the logarithmic number of data points (Log N).
The horizontal dotted
line represents the observed $\langle M_{12}\rangle$ of the Bullet Cluster \citep[$\langle M_{12}\rangle=8.25\times10^{14}\Msun$;][]{Clowe04,Clowe06,Bradac06}.
The dot-dashed lines represent $\langle M_{12}\rangle$ and $\vonetwo$ for the initial condition requirements of \citet{Mast08},
while the dashed lines represent the initial requirements of \citet{Lage14b}.
}
\label{fig:v12}
\end{figure}

We calculate pairwise velocities ($\vonetwo = |\vec{v}_1 - \vec{v}_2|$)
for all halo pairs with $d_{12}\leq10\,\rm{Mpc}$.  To examine how
$\vonetwo$ relates to the mass of the halo pair, we plot the average
halo pair mass ($\langle M_{12} \rangle \equiv (M_{1}+M_{2})/2$)
as a function of $\vonetwo$ in Figure~\ref{fig:v12}, for RS \& FOF
groupings in all three box sizes.  
RS clearly has a broader
distribution along $\vonetwo$ in every case, a direct result of the
code identifying more velocity-space substructure.  

The dotted line denotes the average observed mass of the two components of the Bullet Cluster 
\citep[$\langle M_{12}\rangle=8.25\times10^{14}\Msun$;][]{Clowe04,Clowe06,Bradac06},
dot-dashed lines represent the average mass and pairwise velocities of the initial condition requirements of MB08 ($\langle M_{12}\rangle=4.14\times10^{14}\Msun$ \& $\vonetwo\simeq3000\kms$),
and dashed lines are the initial condition requirements of LF14 ($\langle M_{12}\rangle=1.08\times10^{15}\Msun$ \& $\vonetwo\simeq2799\kms$).
We note that, in our larger volumes, 
RS identifies numerous potential
Bullet progenitor candidates (in and around the upper right quadrant 
of each panel), whereas FOF identifies none.

\subsection{Pairwise velocity probability distribution function}
 \label{sec:pdf}
To estimate the probability of finding a Bullet Cluster candidate,
the approach taken in previous works \citep{Lee10,Thompson12,Bouillot14}
is to identify halo pairs with high relative velocities 
($\vonetwo\geq3000\kms$; MB08)
from among all halo pairs above a given mass threshold 
($\monetwo\geq10^{14}\Msun$)
and separated by $d_{12}\leq10\rm{Mpc}$ .
The Bullet Cluster, however, has a considerably higher mass, with the main cluster having 
an observed mass 
in excess of $10^{15}\Msun$
\citep{Clowe04,Clowe06,Bradac06}.
Given that the usual criteria may not select pairs with mass ratios similar to the Bullet-Cluster, we refer to 
pairs with 
$\monetwo\geq10^{14}\Msun$
as \sampleOne pairs.
To more accurately sample 
potential Bullet-like candidates of higher mass,
we additionally restrict our sample to pairs whose average mass is greater than 
$4.14\times10^{14}\Msun$ (MB08), or $1.08\times10^{15}\Msun$ (LF14) and refer to this sample as \sampleTwo pairs.
In this section we discuss the general pairwise velocity probability distribution function (PDF) of both \sampleOne and \sampleTwo pairs before discussing the number density of these objects in the following section.

Previous works have determined
the probability of finding a \sampleOne pair to be on the order of P($\geq3000\kms)\simeq10^{-8}-10^{-9}$
\citep{Lee10,Thompson12}.  Recently \citet{Bouillot14} argued for
a value two orders of magnitude larger (P$\simeq10^{-6}$) through
the examination of a simulation with a much larger volume 
($29\,\rm{Gpc})^{3}$.
As we show in Section~\ref{sec:v12} however,
identifying haloes with an FOF algorithm can lead to substantially
lower values of $\vonetwo$ (Figure~\ref{fig:v12}) due to its inability
to distinguish between substructure in merging systems like the
Bullet.  These lower values of $\vonetwo$ will have a direct impact
on the resulting probabilities.

\begin{figure}
\includegraphics[scale=0.45]{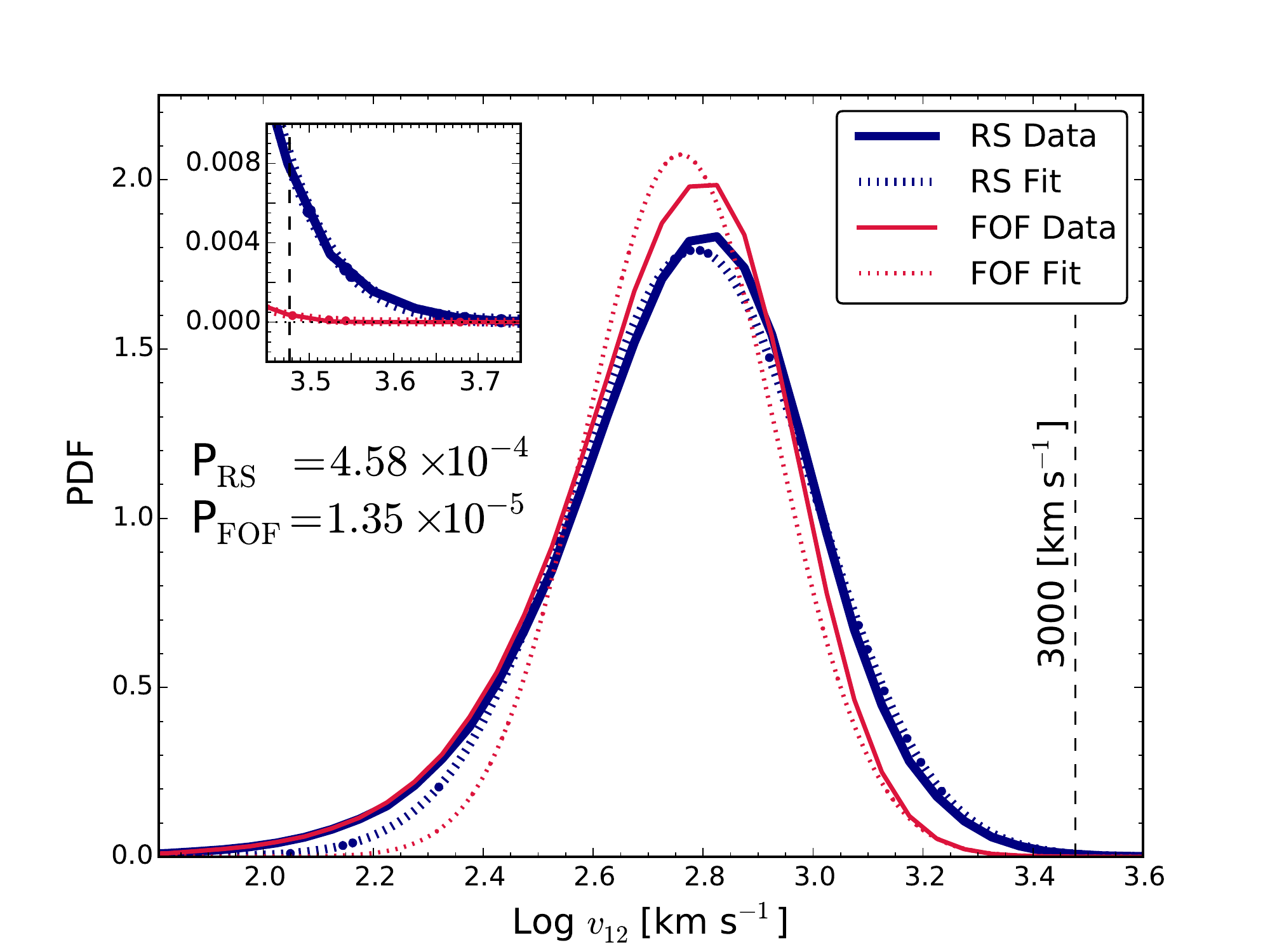}
\caption{
Probability distribution function of massive halo pairs 
($\monetwo\geq10^{14}\Msun, d_{12}\leq10\rm{Mpc}$) 
in our
largest simulation (\BOXTHREE) identified by FOF and RS.  Thin red solid line represents
the FOF data and the thick blue solid line shows RS.  Dashed lines represent
the skew normal fit to the data.  Small error bars were applied at
$\vonetwo\geq 3000\kms$ (MB08) to force the fit to be better there, at
the expense of a poor fit at lower $\vonetwo$.
Inset shows the fit at $\vonetwo\geq3000\kms$,
demonstrating the excellent fit in the high velocity tail.  
We also show the probability of finding a halo pair with $\vonetwo\geq 3000\kms$
obtained by integrating the fitting functions to $\infty$.
}
\label{fig:PDF}
\end{figure}

Figure~\ref{fig:PDF} shows the PDF and fit from our largest simulation
for both RS \& FOF groupings of \sampleOne ($\monetwo\geq10^{14}\Msun,d_{12}\leq10\rm{Mpc}$) pairs.  The overall distribution is
Gaussian-like, with a peak at $\vonetwo\approx 600\kms$.  This is
comparable to the Hubble velocity for haloes separated by 6.5~Mpc.
The \sampleOne pairs in this sample lie in the extreme tail of the distribution
at $\vonetwo\geq 3000\kms$.  Hence such high velocities are only
likely to arise in systems that are merging towards each other.

The probability for a \sampleOne pair is the area under this
curve above the velocity threshold, divided by the total area under
the curve.  Because this high pairwise velocity tail is sampled by a small
number of haloes owing to the limited simulation volume, it may not
be a fair representation of the true statistics to simply count
pairs above this threshold.  One approach to mitigate this is to
fit the PDF with an analytic function and integrate this function
out to infinity.  Previous works have used a Gaussian for this
purpose \citep{Lee10,Thompson12}, but \citet{Bouillot14} argues for
the use of ``Extreme Value Statistics"
\citep[EVS;][]{Frechet27,Fisher28,Gumbel35,Gnedenko43} since the extreme
tail of the PDF can deviate significantly from a Gaussian.

Here we instead follow the approach of fitting a skewed Gaussian,
but we force the fit to be very good particularly for the PDF above
$3000\kms$.  This is accomplished by allowing large fitting errors
at $\vonetwo<3000\kms$, and small ones at $\vonetwo\geq3000\kms$.
A least-square-fit then obtains a very good fit at high velocities,
at the expense of a poorer fit at lower velocities.  However, we
do not need to use the fit at lower velocities, since there we can
directly count pairs within a large and representative sample.
We then calculate the probability of finding a \sampleOne pair
with desired statistics by integrating our best-fit skewed Gaussian
to infinity.  

We also show the skewed Gaussian fits (dashed lines) in Figure~\ref{fig:PDF} .  
At $\vonetwo<3000\kms$ the fit is not good, 
but as shown by the inset in the upper left, 
the fit is much better at $\vonetwo\geq3000\kms$.  
By integrating the fitting function from $3000\kms$ to infinity we obtain
probabilities of P$_{\rm{RS}}=\PRSMB$ \&
P$_{\rm{FOF}}=\PFOFMB$.  Again, this is the probability
of finding a halo pair with $\vonetwo\geq3000\kms$ among all halo pairs with 
$d_{12}\leq10\rm{Mpc}$ and $\monetwo\geq10^{14}\Msun$
at $z=0.489$.

Our P$_{\rm{FOF}}$ is slightly larger than the value calculated by
\citet{Bouillot14} which may be due to our simplified approach.  If
we run the same analysis on FOF groupings with $b=0.20$ as opposed
to $b=0.15$, we find P$_{{\rm FOF},b=0.20}=3.51\times10^{-6}$, which
is slightly smaller than the value of P$=6.4\times10^{-6}$ that
\citet{Bouillot14} obtained using $b=0.15$ for their FOF groupings.
Regardless of this difference, P$_{\rm{RS}}$ remains almost two
orders of magnitude higher than previous estimates using FOF.

Repeating the same exercise on the smaller boxes results in similar
probabilities.  For the \BOXTWO run, we find 
P$_{\rm{RS}}=5.64\times10^{-4}$ and P$_{\rm{FOF}}=4.48\times10^{-6}$.  
And P$_{\rm{RS}}=2.46\times10^{-4}$ and P$_{\rm{FOF}}=3.37\times10^{-6}$ 
for the \BOXONE run.  Note that
the distribution gets noisier with decreasing box size, hence the
probabilities become more unreliable.  Even so, the probabilities
remain roughly similar in order of magnitude, showing that this approach
is stable against reasonable box size variations.

\begin{figure}
\includegraphics[scale=0.45]{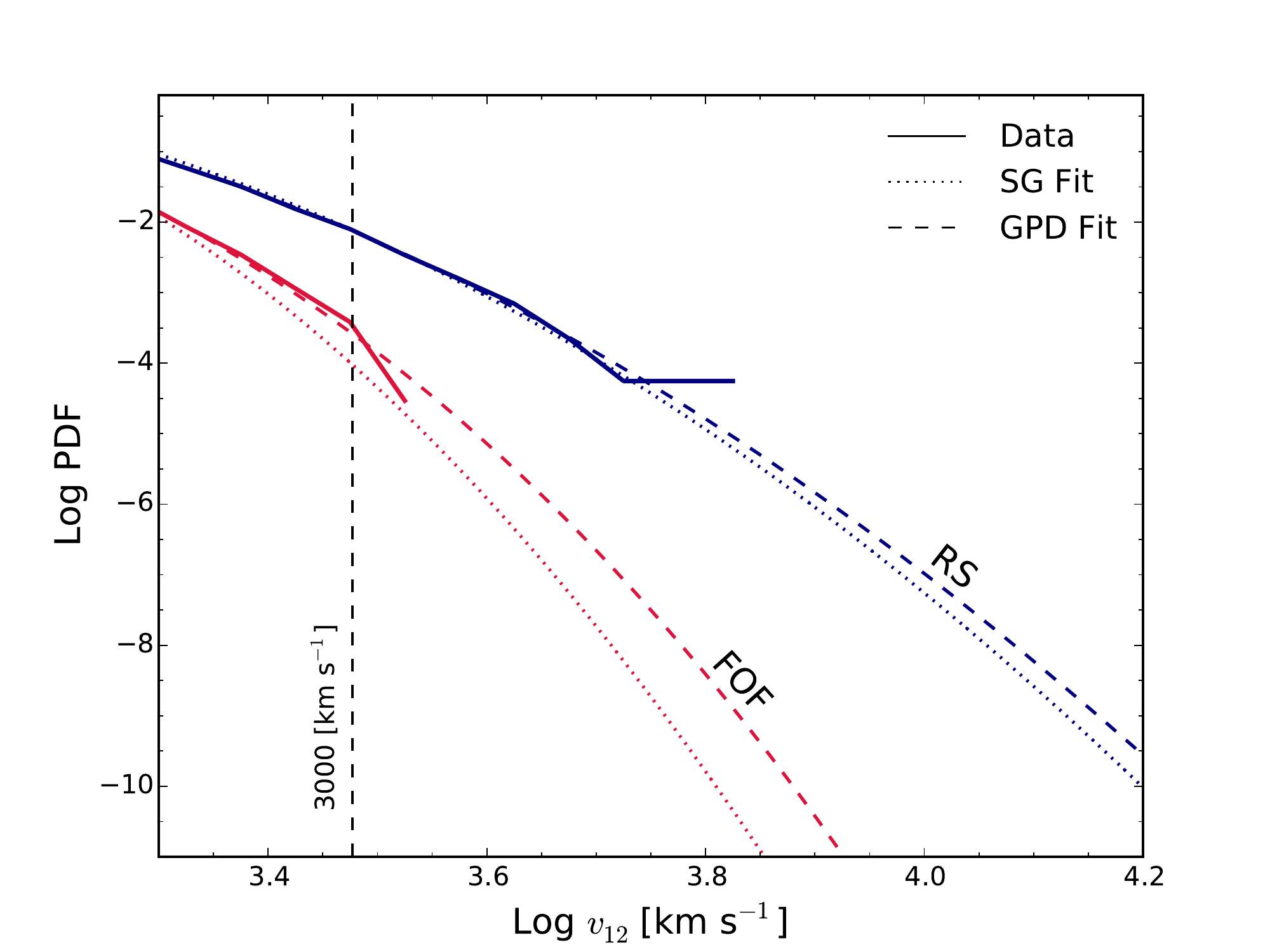}
\caption{
Figure~\ref{fig:PDF} plotted in log-log space, zoomed in on the high pairwise velocity tail.
The skewed normal fits (SG; described in Section~\ref{sec:pdf}) are plotted as the dotted lines, while the Generalized Pareto distributions (GPD) are shown as the dashed lines.
}
\label{fig:GPD}
\end{figure}

If we instead employ EVS and fit a Generalized Pareto distribution (GPD) of the form
\begin{equation}
\rm{PDF} = \frac{1}{\sigma} \left(1+\xi\frac{{\it v}_{12}-\mu}{\sigma}\right)^{1/\xi+1},
\end{equation}
as was done in \citet{Bouillot14}, we find that the probabilities do not differ drastically from our above described method.
We select the location parameter ($\mu$) in the same fashion as \citet{Bouillot14} ($\mu=1600 \kms$; see their Figure~10)\footnote{We note that the exact value of $\mu$ has a  minimal impact on the resulting probabilities} before fitting for the scale ($\sigma$) and shape ($\xi$) parameters.
These fits are shown in Figure~\ref{fig:GPD} for our largest simulation as the dashed lines.
Integrating these from 3000 $\kms$ to infinity we find P$_{\rm{RS,GPD}}=4.51\times10^{-4}$ and P$_{\rm{FOF,GPD}}=9.31\times10^{-6}$, which are different by a factor of 1.02 and 1.45 when compared to our skewed normal fits.
These minor differences indicate that our method is sufficient at fitting the high-$\vonetwo$ tail of the distribution.

\begin{figure}
\includegraphics[scale=0.45]{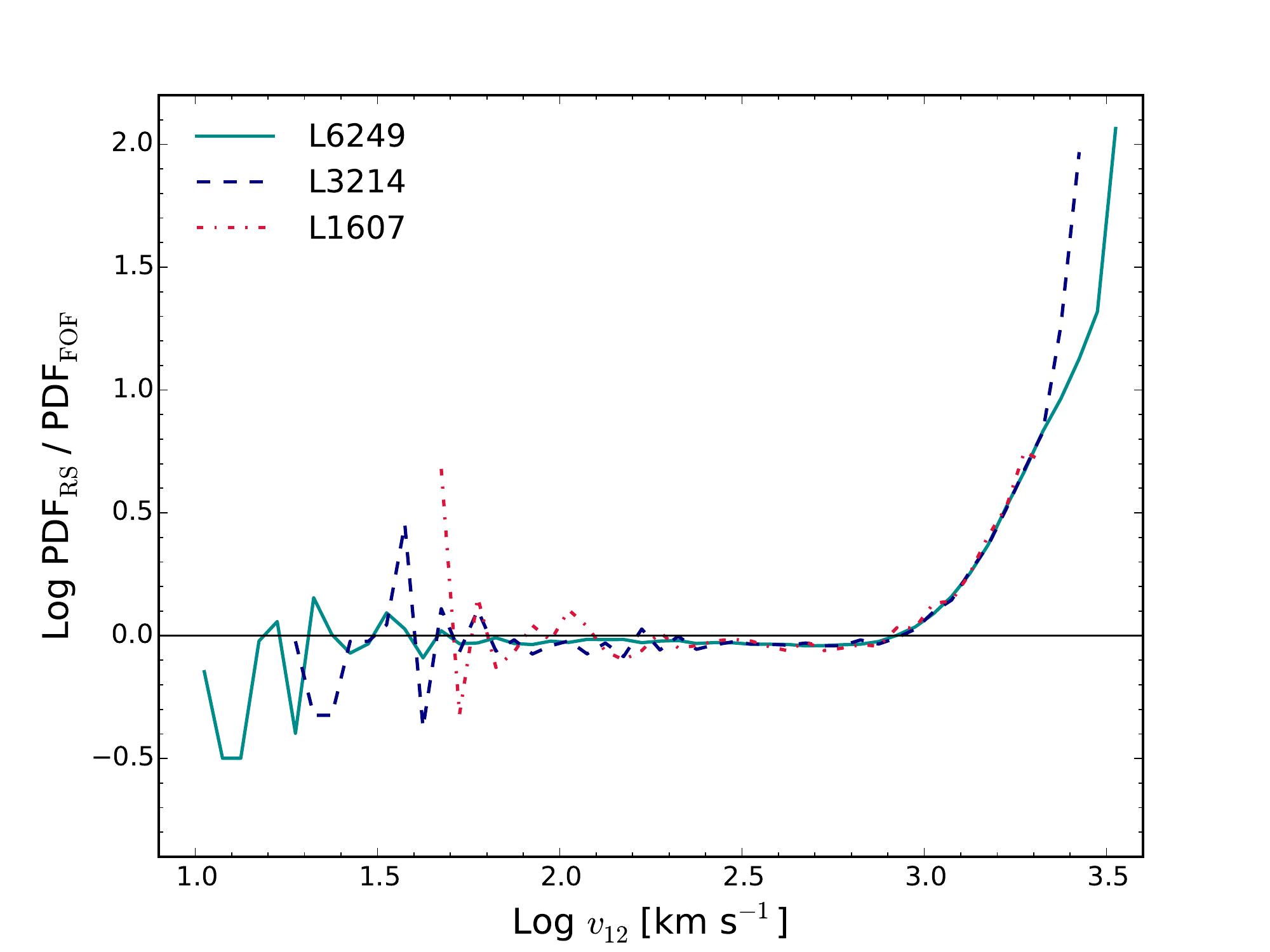}
\caption{Ratio between FOF \& RS PDFs, for all three simulation box sizes.
The two halo finders agree well up to $\vonetwo\sim 1000\kms$, but
RS rapidly increases relative to FOF above this.  The PDF are consistent
among the various box sizes.}
\label{fig:PDFdiff}
\end{figure}

In Figure~\ref{fig:PDFdiff}, we show the ratio between the RS and
FOF PDFs in our three simulation volumes using the skewed normal fits.  
The PDFs are very similar
for $\vonetwo\la 1000\kms$, but above this value the RS probability
increases markedly relative to FOF, such that by $\vonetwo\sim
3000\kms$ it is two orders of magnitude higher.  The ratio of RS
\& FOF PDFs in the \BOXTWO and \BOXONE runs have the same trend as the
\BOXTHREE run, suggesting that the statistical relation between massive
RS \& FOF velocity pairs does not vary drastically even in volumes
as small as 
$(1607\,{\rm Mpc})^3$.

Using the methods described above, we impose an additional mass criteria of 
$\left<M_{12}\right>\geq4.14\times10^{14}\Msun$ (MB08)
and calculate the probability of finding a \sampleTwo halo pair 
within our largest volume to be 
P$_{{\rm RS},Bullet-like}=\BLPRSMB$ and P$_{{\rm FOF},Bullet-like}=\BLPFOFMB$.  
Note that with the additional mass cuts we are sampling a different population of halo pairs.
This results in a value of P$_{{\rm RS},Bullet-like}$ that is
$\approx3\times$ larger than P$_{\rm{RS}}$, 
while P$_{{\rm FOF},Bullet-like}$ is $\approx2\times$ smaller than P$_{\rm{FOF}}$,
indicating that a greater fraction of $\left<\rm{M}_{12}\right>\geq4.14\times10^{14}\Msun$ 
halo pairs within the RS groupings have a $\vonetwo$ greater than $3000\kms$.
Nonetheless, we will show in the next section that such \sampleTwo pairs are globally less frequent than \sampleOne pairs by an order of magnitude.

We note that for the above quoted statistics we are imposing the velocity requirement of $\vonetwo\geq3000\kms$ from MB08 in order to compare to previous works.
If we instead use the velocity requirement of $\vonetwo\geq2799\kms$ argued by LF14 
we find 
P$_{\rm{RS}}(\geq2799\kms)=\PRSLF$ and P$_{\rm{FOF}}(\geq2799\kms)=\PFOFLF$
amongst all halos with $\monetwo\geq10^{14}\Msun$.
Imposing the additional mass cut of $\left<M_{12}\right>\geq1.08\times10^{15}\Msun$ (LF14) we find 
P$_{{\rm RS},Bullet-like}(\geq2799\kms)=\BLPRSLF$ and P$_{{\rm FOF},Bullet-like}(\geq2799\kms)=\BLPFOFLF$.
A summary of these probabilities can be found in Table~\ref{table:probabilities}.

\subsection{Number density estimation}
\label{sec:ndensity}
In order to quantify whether these objects are a likely occurrence in $\lcdm$, 
we must estimate their number densities.
This will also provide predictions for future all-sky surveys that may
be able to probe deeper and hence see such objects over a larger
volume.

In our largest volume simulation, the number of halo pairs with
$d_{12}\leq10\rm{Mpc}$, $\monetwo\geq10^{14}\Msun$, and
$\vonetwo\geq3000\kms$ 
(MB08)
is 6 using FOF and 318 using RS.  To obtain
the full number density, we must additionally correct for our finite volume.
Using the integrated PDF (P) described in previous sections, the number of halo pairs within our sample above the $\vonetwo$ threshold ($N(\geq v_{th})$), and the total number of halo pairs within our sample ($N$; Section~\ref{sec:pdf}) we calculate the correction factor to be
\begin{equation}
n_{correction} = \frac{\rm{P}}{N(\geq v_{th})/N}.
\end{equation}
This results in an additional factor of $1.03$ for RS and $1.53$ for FOF.  
Dividing by our simulation volume, we thus obtain the values of
$n_{\rm{RS}}=\nRSMB\rm{Mpc}^{-3}$
and
$n_{\rm{FOF}}=\nFOFMB\rm{Mpc}^{-3}$.

We note that \citet{Thompson12} computed a number density, but to
do so they needed to extrapolate their cumulative
$\vonetwo$ distribution out to 3000$\kms$ owing to their limited
volume of 
$(2.8\,\rm{Gpc})^3$.
They calculated a value of
$n_{\rm{FOF}}=9.4\times10^{-13}\rm{Mpc}^{-3}$
for such halo pairs.
The use of a larger volume and a better fitting function results
in significantly higher values for FOF, and RS additionally provides
a number density increase by more than a factor of 30.

Again the distinction must be made that the above values of $n$ 
represent the number density of \sampleOne halo pairs
matching the velocity cut of MB08.
Within our largest volume simulation we find 0 FOF pairs, and 35 RS pairs that meet our \sampleTwo criteria.
We find the number density of \sampleTwo halo pairs to be 
$n_{{\rm RS},massive}=\BLnRSMB$,
i.e. an order of magnitude less than the number density of \sampleOne pairs
($n_{\rm{RS}}$).

Using the velocity criteria set by LF14 we find 15 FOF pairs and 535 RS pairs with $d_{12}\leq10\rm{Mpc}$, $\monetwo\geq10^{14}\Msun$ within our largest simulation.
The number densities of these \sampleOne halo pairs is 
$n_{\rm{RS}}(\geq2799\kms)=\nRSLF$ $\rm{Mpc}^{-3}$ 
and
$n_{\rm{FOF}}(\geq2799\kms)=\nFOFLF$ $\rm{Mpc}^{-3}$.
Again we impose an additional mass cut from LF14 to select \sampleTwo candidates and find a number density of
$n_{{\rm RS},massive}(\geq2799\kms)=\BLnRSLF$ $\rm{Mpc}^{-3}$.
Table~\ref{table:probabilities} summarizes these results.

\subsection{Bullet-like pair candidates}
\label{sec:candidates}

We now study in more detail the properties of \sampleTwo pairs
in the simulation.  \SampleTwo 
pairs are selected from
our largest simulation according to the
MB08 criteria:
(i) $\monetwo\geq10^{14}\Msun$,
(ii) $\left<\rm{M_{12}}\right>$$\geq4.14\times10^{14}\Msun$,
(iii) $d_{12}\leq10\,\rm{Mpc}$ and
(iv) $\vonetwo\geq3000\kms$.  
As mentioned in Section~\ref{sec:ndensity} we find 
35
candidates
within the RS groupings that meet this criteria, and zero within
the FOF groupings.

\begin{figure*}
\includegraphics[scale=0.5]{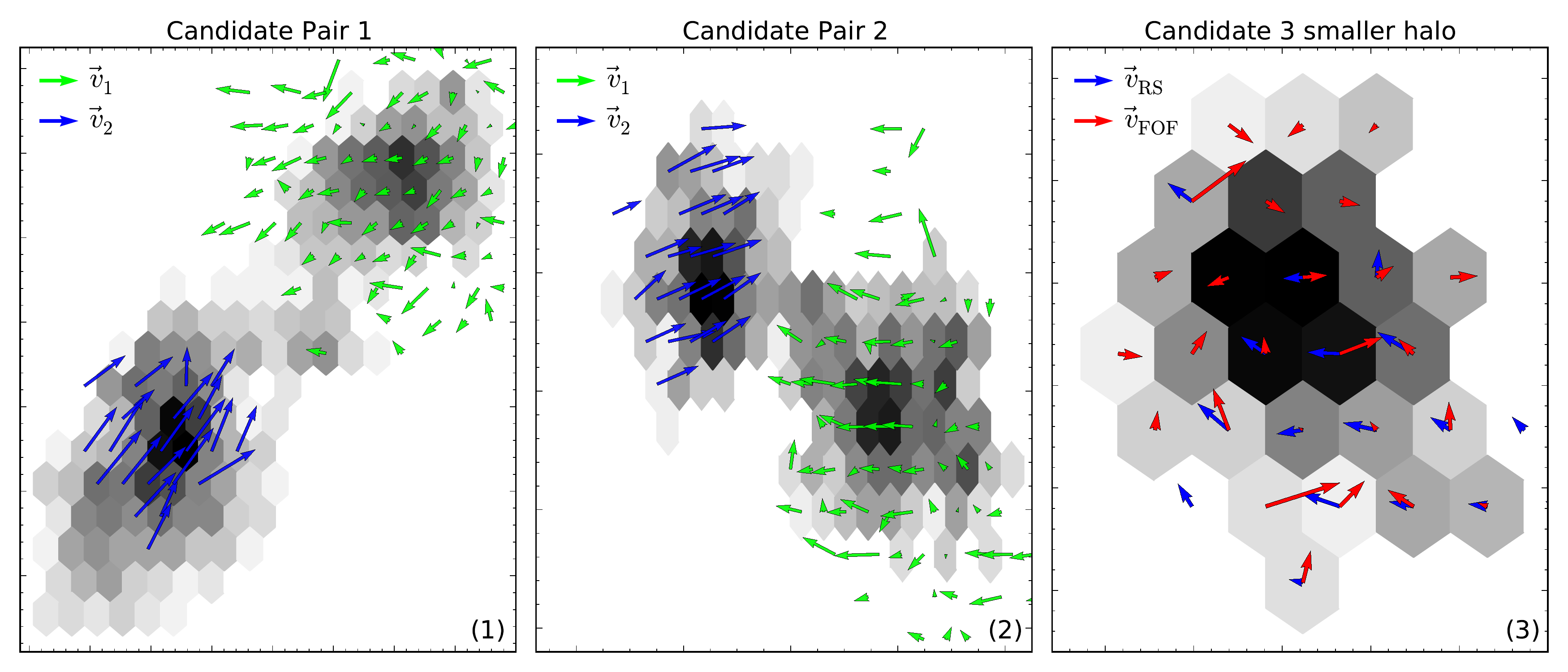}
\caption{Candidate halo pairs from Table~\ref{table:candidates} projected on the $x-y$ plane.  Greyscale
intensity represents the number of FOF identified particles contained
within each hexagonal bin.  The size of each bin corresponds roughly
to the FOF linking length of $\simeq0.4\rm{Mpc}$ ($b=0.15$).  Panels
(1) \& (2) show RS-identified mean particle velocity vectors from
each bin.  To accentuate the differences we also subtract the FOF
halo's bulk velocity from the RS velocities.  In these cases,
FOF groups the two concentrations into a single halo, whereas
RS separates them into two haloes based on their distinct kinematics.
Panel (3) only shows
the smaller halo from candidate pair \#3; here we also overplot the FOF
velocity vectors in red.  Where most of the particles
lie (the dark region), the velocity vectors of RS and FOF are
in the opposite direction, showing that again FOF is merging kinematically
distinct components that RS separates.  The resulting pairwise velocity
relative to the main halo (not shown) is much smaller in the FOF case.
}
\label{fig:vis}
\end{figure*}

We select three ideal Bullet-like candidate pairs from the 
RS \sampleTwo sample
that best match the Bullet Cluster mass ratio and
separation distance, and examine FOF data for the same haloes.  
The results are summarized in Table~\ref{table:candidates}.  RS Bullet-like
candidates \#1 and \#2 are each grouped into a single FOF halo.
Pair three however, is identified as two separate FOF groups, but
with a $\vonetwo$ that is less than half of its RS counterpart (and
hence fails to meet the \sampleTwo criteria).  We note that we
used $b=0.15$ to group FOF haloes; if we had used the canonical value
of $b=0.20$, FOF identifies system \#3 as a single group.

By visualizing these systems we can better understand the differences
between RS and FOF.  In Figure~\ref{fig:vis} we project the halo
particles of our candidate groups onto the $x-y$ plane, and then
bin them into hexagonal bins\footnote{Interactive 3D visualizations are available at \url{http://www.physics.unlv.edu/~rthompson/bulletCandidates}}
\footnote{Hexagonal bins are the closest one can get to a circle while still allowing the shapes to interlock.  This makes a hexagonal tessellation the most efficient and compact division of 2D data.}.  
The number of FOF-identified particles within a given
bin is indicated by the shade of each hexagon, with darker shades
corresponding to more particles contained within.  Additionally,
we indicate the mean velocity vectors of particles within each bin
by the colored arrows.

Panels (1) \& (2) show halo candidate pairs \#1 and \#2 from
Table~\ref{table:candidates}.  Both of these pairs are identified
as a single group by FOF; one can clearly see the `bridge' of
particles connecting the two concentrations.  Arrows indicate the
mean velocity of the RS-identified particles within a given bin
minus the bulk velocity of the corresponding (single) FOF halo.
The directions and magnitudes of the different RS groupings provides
clear evidence that the single object identified by FOF is indeed
two separate objects when viewed in velocity space.  We do not 
show FOF velocity information, but it is very similar to that of
RS since the actual particles grouped into RS and FOF haloes are
quite similar; the difference is that, by using velocity information,
RS is able to separate these systems into two distinct haloes whereas
FOF (which does not use velocity information) lumps them into one.

Candidate \#3 is distinct since both FOF and RS identify them as
two separate haloes.  Here we only show the smaller `bullet' from
this pair in Panel~(3) of Figure~\ref{fig:vis}, since the larger
halo is identified similarly by both.  We further show the median velocity
vectors for both FOF-identified particles (red) and RS-identified
particles (blue).  While some bins have similar mean velocities,
others are considerably different from one another.  Most notably,
two of the dark center bins where the majority of the mass lies
show nearly opposing velocity vectors between FOF and RS.  What has
happened is that there are multiple objects in this region, and FOF
has overgrouped them resulting in a velocity much closer to that
of the main halo (not shown).  Meanwhile, RS is able to distinguish
the relatively small `bullet' that is distinct in velocity space.
The overgrouping results in an FOF halo that is $7\times$ more
massive than the RS counterpart, and double in radius.  Crucially,
the pairwise velocity is reduced by a factor of two when compared
to the RS results.

By lumping together multiple groups into a single object, the overall bulk
velocity can easily get washed out.  Consider this simple example:
two head-on merging haloes are grouped together; their bulk
velocities would effectively cancel out leading to a much lower bulk
velocity for the final group.  When $\vonetwo$ is calculated between
this group and others, the resulting value would be much lower than if they were considered as separate objects.
This problem is exacerbated as the number of distinct objects grouped
together increases.  
Candidate \#3 is one clear example of such a process.

\section{Conclusions}
\label{sec:conclusions}

In this work we determine the probability and number density of
finding systems analogous to progenitors of the Bullet Cluster
1E0657-56 within large-volume cosmological N-body simulations.  We
particularly examine the difference between two popular halo finding
algorithms in the context of searching for a massive, high-pairwise
velocity halo pair.  Our results show that halo finders that only
consider particle positions (FOF) can underestimate the probability
of high pairwise velocity systems, which can ultimately lead to
tension with the $\lcdm$ model.  Halo finders that additionally
consider particle velocities can more robustly
identify kinematically distinct substructures, resulting in greater
$\vonetwo$ probabilities, alleviating tensions with $\lcdm$.

Within our largest cosmological N-body simulation, we find the
probability of producing a halo pair with $\vonetwo\geq3000\kms$ 
\citep{Mast08}
from among all halo pairs with $d_{12}\leq10\rm{Mpc}$, \& $\monetwo\geq10^{14}\Msun$ to be
P$_{\rm{RS}}=\PRSMB$ when using \RS\ (RS).
This value is larger by 1.5 dex than when one only considers particle
positions using a friends-of-friends (FOF) halo finding algorithm 
(P$_{\rm{FOF}}=\PFOFMB$).  Simulation box size
still plays an important role as we show in Figure~\ref{fig:v12},
but using RS, a box size on the order of 
$(3\,\rm{Gpc})^3$
yields similar probabilities as a 
$(6.4\,\rm{Gpc})^3$
box with reasonable
extrapolations of the $\vonetwo$ probability distribution functions.
If we slide the velocity cut back to $\vonetwo\geq2799\kms$ as argued by \citet{Lage14b} we find slightly higher probabilities of
P$_{\rm{RS}}(\geq2799\kms)=\PRSLF$ and
P$_{\rm{FOF}}(\geq2799\kms)=\PFOFLF$.
We also find that these results do not change drastically when fitting with "Extreme Value Statistics" rather than a skewed Gaussian.

We estimate the expected number density of such \sampleOne
objects to be 
$n_{\rm{RS}}=\nRSMB$ $\rm{Mpc}^{-3}$
using the velocity cuts of MB08, and 
$n_{\rm{RS}}(\geq2799\kms)=\nRSLF$ $\rm{Mpc}^{-3}$ 
for LF14.
Imposing an additional mass criteria of 
$\langle \rm{M}_{12} \rangle \geq 4.14\times10^{14}\Msun$ (MB08) 
in order to identify more massive systems
that are truer analogs of the Bullet Cluster, we calculate the
expected number density of \sampleTwo objects to be
$n_{{\rm RS},massive}=\BLnRSMB$ $\rm{Mpc}^{-3}$. 
Including this more stringent mass criterion, RS identifies $\simeq35$
\sampleTwo candidates within our largest simulation, wherein
FOF applied to the same simulation identifies none.
Imposing higher mass cut of $\langle \rm{M}_{12} \rangle \geq 1.08\times10^{15}\Msun$ (LF14)
yields a number density of \sampleTwo objects to be slightly smaller 
($n_{{\rm RS},massive}(\geq2799\kms)=\BLnRSLF$ $\rm{Mpc}^{-3}$). 

By studying individual examples, we show that the differences between
RS and FOF owe to the identification of more substructure by
considering particle velocities.  We identify three ideal candidate
halo pairs from the RS dataset and examine the FOF data in the same
region (Table~\ref{table:candidates}).  By not considering particle
velocities, FOF tends to over-group haloes and/or group together
particles that are clearly different groups in velocity space
(Figure~\ref{fig:vis}).

We do not expect to find an exact match to the Bullet Cluster within
one random realization of our Universe.  The more significant point
is that producing such {\it massive high-$\vonetwo$} pairs should no
longer be considered a challenge to $\lcdm$, as was suggested in
\citet{Lee10} and \citet{Thompson12}.  
As we have shown here, the identification
of such a pair is not only possible but likely when a kinematic
halo finding algorithm is used.  While for the overall halo population
the differences between RS and FOF are fairly minor, using particle
velocity information is crucial when identifying haloes in the context
of this particular problem.

More broadly, this greatly ameliorates a major challenge to the
$\lcdm$ model presented by the high progenitor pairwise velocities
of the Bullet Cluster.  Instead, we show that the Bullet Cluster
is a rare but expected object in a $\lcdm$ universe.  Future all-sky
X-ray surveys 
(e.g., eROSITA\footnote{\tt http://www.mpe.mpg.de/eROSITA})
together with upcoming weak lensing surveys 
(e.g., LSST\footnote{\tt http://www.lsst.org/lsst}) 
will potentially identify many more
Bullet-like systems to lower masses and/or higher redshifts, which
can be used to further explore the nature of dark matter, and
thus test the $\lcdm$ paradigm in more detail.  At this time,
however, the Bullet Cluster provides unequivocal support for the
modern concordance cosmological paradigm.

 \section*{Acknowledgements}
We thank the anonymous referee for their helpful comments.
The simulations used in this paper were run on `Blue Waters' at the
National Center for Supercomputing Applications (NCSA), and the data was
analyzed on the `Timon' cluster at the University of the Western Cape with 
{\small pyGadgetReader}\footnote{\url{https://bitbucket.org/rthompson/pygadgetreader}}\citep{pyGadgetReader} \& 
{\small SPHGR}\footnote{\url{https://bitbucket.org/rthompson/sphgr}}\citep{SPHGR}.
This research is part of the Blue Waters sustained-petascale
computing project, which is supported by the National Science
Foundation (awards OCI-0725070 and ACI-1238993) and the state of
Illinois.  Blue Waters is a joint effort of the University of
Illinois at Urbana-Champaign and NCSA.  This work is also part of
the PRAC allocation, supported by the National Science Foundation
(award number OCI-0832614).  Support for this work was provided by
the South African Research Chairs Initiative and the South African
National Research Foundation, along with NASA grant NNX12AH86G.
KN is grateful to the discussion with Dr. Tomoaki Ishiyama on large N-body simulations.

\begin{onecolumn}


\begin{deluxetable}{rrccccc}
\tabletypesize{\footnotesize}
\tablecolumns{7}
\tablewidth{0pc}
\tablecaption{Probabilities \& number densities}
\tablehead{
	\colhead{Pairs} &
	\colhead{N$_{\rm{RS}}$} &
	\colhead{N$_{\rm{FOF}}$} &
	\colhead{P$_{\rm{RS}}$} &
	\colhead{P$_{\rm{FOF}}$} &
	\colhead{$n_{\rm{RS}}$} &
	\colhead{$n_{\rm{FOF}}$}
}
\startdata
\sidehead {\citet{Mast08} criteria}
\SampleOne & 318 & 6 & $\PRSMB$ & $\PFOFMB$ & $\nRSMB$ & $\nFOFMB$ \\
\SampleTwo & 35 & 0 & $\BLPRSMB$ & $\BLPFOFMB$ & $\BLnRSMB$ & - \\
\hline
\sidehead {\citet{Lage14b} criteria}
\SampleOne & 535 & 15 & $\PRSLF$ & $\PFOFLF$ & $\nRSLF$ & $\nFOFLF$ \\
\SampleTwo & 23 & 0 & $\BLPRSLF$ & $\BLPFOFLF$ & $\,\,\BLnRSLF$ & - \\
\enddata
\vspace{-0.5cm}
\label{table:probabilities}
\tablecomments{
\SampleOne pairs are defined as halo pairs 
with $\vonetwo\geq3000\kms$ (MB08), or $\vonetwo\geq2799\kms$ (LF14)
from among all halo pairs with $\monetwo\geq10^{14}\Msun$ and $d_{12}\leq10\,\rm{Mpc}$.  
\SampleTwo pairs are defined by imposing an additional mass criteria of 
$\left<M_{12}\right>\geq4.14\times10^{14}\Msun$ (MB08) or
$\left<M_{12}\right>\geq1.08\times10^{15}\Msun$ (LF14).
All values are from our largest volume simulation.
`N' is the total number of pairs that meet said criteria, `P' represents the probability (Section~\ref{sec:pdf}), and `$n$' is the number density given in units of 
$\rm{Mpc}^{-3}$.
Previous works found P$_{\rm{FOF}}=6.4\times10^{-6}$ \citep{Bouillot14}, and 
$n_{\rm{FOF}}=9.4\times10^{-13}$ 
\citep{Thompson12} for \sampleOne pairs
when imposing the MB08 velocity cut.
}
\end{deluxetable}

\begin{deluxetable}{ccccccccc}
\tabletypesize{\footnotesize}
\tablecolumns{9}
\tablewidth{0pc}
\tablecaption{Bullet-like candidate pairs}
\tablehead{\multicolumn{1}{c}{Pair} &
	\colhead{$\vonetwo$} &
	\colhead{$d_{12}$} &
	\colhead{$\theta$} &
	\colhead{M$_1$} &
	\colhead{M$_2$} &
	\colhead{Mass} &
	\colhead{r$_{1}$} &
	\colhead{r$_{2}$} \\
	\colhead{ \ } &
	\colhead{[km 
	s$^{-1}$]} &
	\colhead{[Mpc]} &
	\colhead{ \ } &
	\colhead{[$\Msun$]} &
	\colhead{[$\Msun$]} &
	\colhead{Ratio} &
	\colhead{[Mpc]} &
	\colhead{[Mpc]} 
}
\startdata
\sidehead{Rockstar Candidates}
1 & 4893 & 5.95 & 156 & 1.69e15 & 2.01e14 & 0.08  & 2.34 & 1.00 \\
2 & 3506 & 3.57 & 149 & 1.65e15 & 2.06e14 & 0.13  & 2.32 & 1.16 \\
3 & 3130 & 6.17 & 141 & 1.88e15 & 1.30e14 & 0.07  & 2.42 & 0.99 \\
\sidehead{FOF Findings}
1 & Single &- & - & 4.19e15 & - & - & 3.17 & - \\
2 & Single & - & - & 2.48e15 & - & - & 2.66 & - \\
3 & 1537 & 5.69 & 131 & 1.56e15 & 9.21e14 & 0.59 & 2.28 & 1.91 \\
\enddata
\vspace{-0.5cm}
\label{table:candidates}
\tablecomments{
Selected \sampleTwo candidate pairs from our largest simulation (see Table~\ref{table:sims}).
Rockstar candidates were chosen based on how similar they were to the Bullet Cluster initial condition requirements of \citet{Mast08}.
Corresponding FOF haloes were then identified based on their proximity to the chosen RS haloes.
}
\end{deluxetable}

\end{onecolumn}
\end{document}